\documentclass[prl,twocolumn,showpacs,floatfix]{revtex4-1}
\usepackage{psfrag}
\usepackage{amssymb,amsmath,amsthm}
\usepackage[dvips]{graphicx}
\usepackage{verbatim}
\usepackage[colorlinks=true,linkcolor=blue,citecolor=blue]{hyperref}

\begin{document}

\title{Strong interaction between plants induces circular barren patches: fairy circles}
\date{\today }
\author{C. Fernandez-Oto$^{1}$, M. Tlidi$^{1}$, D. Escaff$^{2}$ and M.G. Clerc$^{3}$}
\affiliation{$^{1}$Facult\'e des Sciences, Universit\'{e} Libre de
Bruxelles (U.L.B.), CP\ 231, Campus Plaine, B-1050 Bruxelles, Belgium}
\affiliation{$^{2}$Complex Systems Group, Facultad de Ingenier\'\i a y
Ciencias Aplicadas, Universidad de los Andes, Av. San Carlos de
Apoquindo 2200, Santiago, Chile}
\affiliation{$^{3}$Departamento de F\'isica, Universidad de Chile, Blanco Encalada 2008, Santiago, Chile}
\begin{abstract}
Vast landscapes extending from southern Angola, Namibia, and South Africa exhibit localised barren patches of 
vegetation  called  fairy circles. They consist of isolated or randomly distributed circular areas devoid of any 
vegetation. They are  surrounded by fringes of tall grasses and are embedded in a sparse grassland on sandy 
soils. When the aridity increases, the size of the fairy circles increases and 
can reach diameters up to 10m.  Although several hypotheses have been proposed,  the origin 
of this phenomenon remains unsolved.   We show that
a simple non-local model of plant ecology based on realistic assumptions provides a quantitative
explanation of this phenomenon. Fairy circles result from strong interaction between interfaces connecting two uniform covers: 
the uniform  grassland and the bare states, and their stabilisation is attributed to the Lorentzian-like 
non-local coupling that models the competition between plants. The cause of their formation in thus rather inherent to the vegetation dynamics. Our analysis explain how a circular 
shape and fringes are formed,  and how the aridity level influences the size of the fairy circles. In agreement 
with field observation, these theoretical findings, provide a link between a strong non-local interaction 
of plants and the formation  of stable fairy circles.  Finally, we show that the proposed mechanism is model-independent: indeed, it also applies to 
the reaction-diffusion type of model that emphasises the influence of water transport on the vegetation dynamics.
\end{abstract}

\maketitle
\begin{figure}[t]
\centerline{\includegraphics[width=8cm]{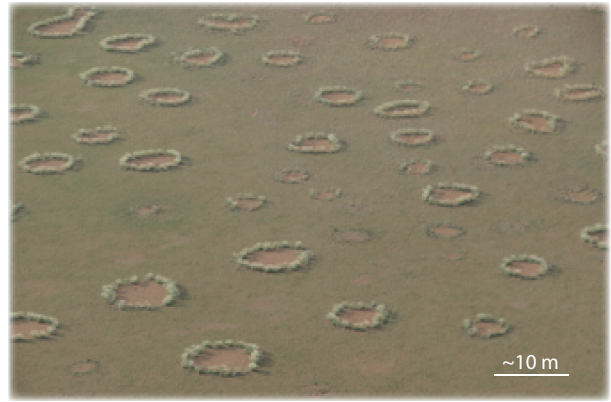}}
\caption{Typical random distribution of fairy circles observed in the 
pro-Namib zone of the west coast of Southern Africa (courtesy of M. Johny Vergeer).}
\label{TW1}
\end{figure}

When ecosystems are subject to limited resources such as water and nutrients, they 
adopt a periodic \cite{LL}-\cite{Sherat} 
or aperiodic \cite{LTC}-\cite{Meron2} distribution of densely vegetated and  bare soil areas.  
In order to fight against drought, every plant struggles to spread its roots so 
that they outgrow the size of the aerial structure  by an order of magnitude, for greater water uptake. However, this adaptation  increases plant-to-plant competition between neighbouring 
plants, and at plant communities level, via non-local interaction,  favours the self-organisation 
phenomenon \cite{LL}.   Other modelling approaches that underline either the role of water transport by below ground and above ground run-off  \cite{Meron1}, or the role of constructive influence of the environment randomness \cite{Ridolfi} have been proposed to explain the formation of vegetation patterns.   A well-documented example of non-uniform spatial distribution of biomass is localised
barren patches of vegetation often called \textit{fairy circles} (FCs), but their origin is still a subject of debate. These circles can be either 
isolated, or randomly distributed in space. They are embedded in a sparse grassland as 
shown in   Fig.~\ref{TW1}. The term
"grass" refers to any grass or herbaceous species. This phenomenon is visible from either aerial and satellite photographs or on ground level in vast territories in southern Angola, Namibia, and South Africa \cite{Getzin,Albrecht}. In these arid territories, the annual rainfall ranges between $50$ and $100$~mm. The size of barren patches increases from South to North where the climate \cite{Climate} becomes more and more arid. Their average diameter ranges from $2$~m to $10$~m \cite{Rooyen}.  
Several hypothesis attributing the formation of FCs to factors external to the vegetation have been suggested \cite{Getzin}-\cite{Jorgen}.
All these external factors can not explain the orgin of the circular shape of FCs \cite{Rooyen,Grube}. Recently,  a detailed description of fairy circle structures, of their life span and  range of variation in multiple habitats  has been provided \cite{Walter}.  How Mathematical modelling of the vegetation growth constitutes an important tool toward the understanding of arid ecosystems. To explain the origin of this ecological phenomenon, 
we take a strictly homogeneous (flat) territory and isotropic environmental conditions.  This 
corresponds to a reasonable approximation for a large territory compared  to typical sizes of FCs and for a small territory compared to the geographical scale. We use the generic interaction-redistribution 
model \cite{LL}.  When the water resources are scarce,  in order to survive plants increase the size of their roots structure. The size of the roots in dryland can reach ten time the size of 
the aerial structure (see Fig. 7 of Ref.~\cite{Albrecht}). In this arid climatic 
condition, plants should then compete for the extraction of water. 

The purpose of this Letter is to report on the occurrence of stable FCs with an intrinsic dynamical nature. The size of the FC is determined by the strong 
non-local competitive interaction between plants mediated by a Lorentzian type of kernel function, 
and under bistability between spatially homogeneous covers of arid ecosystems. The size of the FCs is thus determined rather by the system's dynamical parameters and not by external factors and/or boundary conditions. In agreement 
with field observations, we show that the diameter of a FC increases with the aridity parameter, 
and each isolated FC  exhibits a fringe with high biomass density. In order to show that our mechanism is  model-independent,  we incorporate the  Lorentzian-like non-local coupling in the reaction-diffusion type  of model that includes water transport \cite{Meron1}, and we also show that  this model supports stable FCs. Finally, we show that when FCs diameter exceeds  a given maximum size, {\it i.e.,} maximum aridity, they   present a deformation of their circular shape. However, when the aridity is  lower than  a given threshold, fairy circles shrink and disappear. 
\begin{figure}[t]
\centerline{\includegraphics[width=8cm]{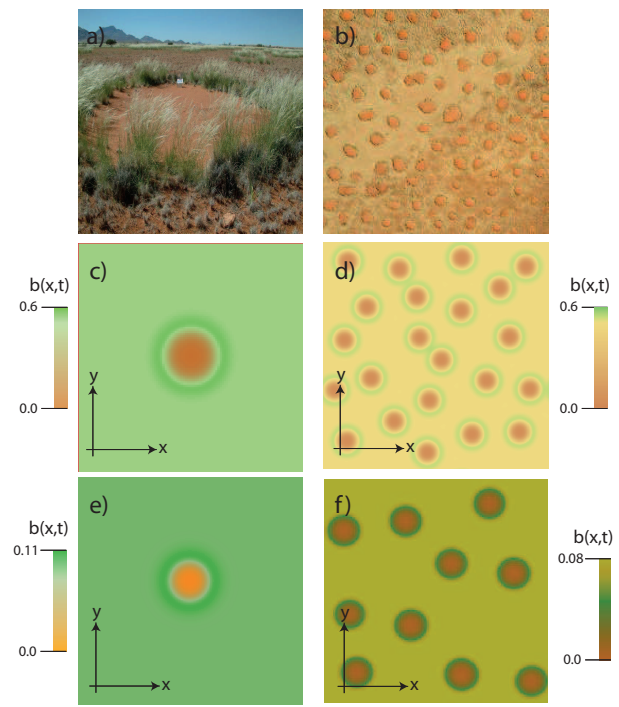}}
\caption{(a,b) Typical fairy circles observed in the pro-Namibia zone of the west coast of Southern Africa. (c,d) Snapshots of biomass density obtained by numerical simulations of the model Eq (1) on a $110 \times 90$ for (c) and and $256 \times 256$ for (d). Green colors indicate a uniformly vegetated state of high density, orange colors indicate bare state with zero density. Parameters are $n=2$, $L_c=1.5$, $\xi_c=1.2$, $\xi_f=3$, $\mu=1.23$.  (e,f) Snapshots obtained by numerical simulations of the model  [ Eqs. (1) in Ref \cite{Meron1}] with the parameters $\mu=0.4143$, $\delta_b=\delta_w=1$, $\rho=6$, $\beta=5$, $p=0.29$, $\alpha=10$, $q=0.05$, $f=0.1$, $L=1$ and $n=1.5$.  }
\label{TW2}
\end{figure}
The spatial distribution of all plants is described by  the vegetation density $b=b({\bf {r}},t)$ at time $t$ and at the point ${\bf {r}}=(x,y)$. It is defined as the plant biomass per unit area. We consider a single dominant species forming a plant community that  occupies a flat territory \cite{Greig-Smith,LL}. The growth and  death processes are modelled by the following logistic equation governing the time-space evolution of the biomass \cite{Leferver09}
\begin{equation}
\partial_t b=b\left[1-b\right]\mathcal{M}_f-\mu b\mathcal{M}_c +\nabla ^{2} b.
\end{equation}
The first term  expresses the rate at which the biomass density increases and saturates. This is  the biomass 
gain that  corresponds to the natural production of  plants via seed production, dissemination, 
germination and development of shoots into new mature plants.  The second term models the 
biomass losses which describes death or destruction by grazing, fire, termites, or herbivores. 
The parameter $\mu$ measures the resources scarcity,  {\it i.e.}, the aridity parameter. The Laplacian $\nabla ^{2} b$  expresses the vegetation spatial 
propagation through seed dispersion, production and germination, which we assume to be a simple 
random walk or brownian motion. When the water becomes scarce, plants adapt their roots systems to fight against water scarcity. They strive to maintain their water  uptake by increasing the length of their roots.They thus compete with other plants on long distance $L_c$. This is a negative feedback that tends to reduce the biomass density modelled by the function $\mathcal{M}_c$. In contrast with previous mathematical models of plant ecology, we incorporate a strong non-local  interaction by using a Lorentzian-like kernel. The term "strong" refers to the class of spatial kernel functions that decreases slower for large distance $\bf r^{{\prime}}$ than the exponential distribution \cite{Escaff}. The major ingredient that needs to be incorporated in Eq.~(\ref{TW1}) is the competition by a Lorentzian-like kernel $\mathcal{M}_c$. This function reads 
\begin{equation}
\mathcal{M}_{c}=\exp\left[\frac{\xi_{c}}{N_{c}} \int\frac{b({\bf r+r^\prime},t)}{\left(1+\frac{|\bf r^{\prime}-\bf {r}|^2}{L_{c}^2}\right)^n}d\bf r^{{\prime}}\right],
\end{equation}
where $N_c$ is the normalisation coefficient. The facilitative interaction between plants is modelled by the  function $\mathcal{M}_f$ which expresses the positive feedback which favours the vegetation development. We focus on the limit where the length of the facilitation is negligible $L_f\approx 0$. Then the mean field function $\mathcal{M}_f \approx\exp(\xi_f b)$. The parameters ${\xi_{c}}$ and ${\xi_{f}}$ model the interaction strength associated with the competitive and facilitative. In the absence of non-local interaction ($\xi_{c,f}=0$), Eq. (1) becomes the F-KPP equation, that is a paradigm in the study of front propagation \cite{Fisher,KPP}. Note that, at the F-KPP equation level there is no bistability between  stable spatially uniform states. When non-local interaction is included, however, it is possible to find two coexisting spatially uniform stable states: the bare solution corresponding to a territory totally devoid of vegetation $b_0=0$; and a uniformly vegetated state $b_s>0$. This phenomenon has been studied in Ref.\cite{Tli-Lef-Vla,Leferver09} as well as a pattern forming instability.

Considering a regime far from any pattern forming instability, we focus on the parameters where the bistability between the bare and the uniformly vegetated states takes place. 
To generate FCs, we need to connect  these two homogeneous  states.  This connection is refered to as front.  
Depending on the choice of the non-local coupling function, front interaction between vegetated and bare states can be either strong or weak \cite{Escaff}.  All ecol	l models used a weak non-local coupling such as a Gaussian or an  exponential function \cite{LL,Meron1,Meron2}, which decay asymptotically faster than the Lorentzian  distribution. When considering a weak front interaction,  domains of bare state embedded in the uniformly vegetated state are unstable: they either shrink or broaden. This type of non-local coupling envolves short range interaction which is always attractive. However, when considering a strong non-local coupling mediated by a Lorentzian type of function, this leads the stabilisation of FCs. Contrary to weak nonlocal coupling, strong competitive  interaction induces long-range interaction that can be either repulsive or attractive. This allows for a confinement of the barren state domain inside a uniformely vegetated state. There is then a balance between strong competitive  interaction and the tendency of grass to populate the barren zone. Indeed, numerical simulations of Eq. (\ref{TW1}) with periodic boundary conditions allow us to obtain 
stable FCs as shown in Figures~\ref{TW2}b and  ~\ref{TW2}c.   Photos and areal views of the pro-Namibia fairy circles are shown in Figures~\ref{TW2}a,  \ref{TW2}b. These long-lived permanent circular structures neither  shrink in spite of available free space, nor grow  in spite of adverse conditions. In agreement with long-term observations, Van Rooyen et al.  \cite{Rooyen} show that several marked fairy circles are stable more than 20 years later.  In order to show that bistability together with strong non-local coupling are responsible for the stabilisation of fairy circles, we use the reaction-diffusion model  [see Eqs. (1) in Ref \cite{Meron1}], replacing the non-local coupling $g({\bf r}+  {\bf  r^\prime},t)$ by a Lorentzian like function.   This model emphasises the role of water transport  by below ground and/or above ground run-off.  The results of  numerical simulations of this model are illustrated in  Figs.~ \ref{TW2}e and  ~\ref{TW2}f. These results  show  that the formation of stable FCs are model-independent. 
\begin{figure}[t]
\centerline{\includegraphics[width=8.cm]{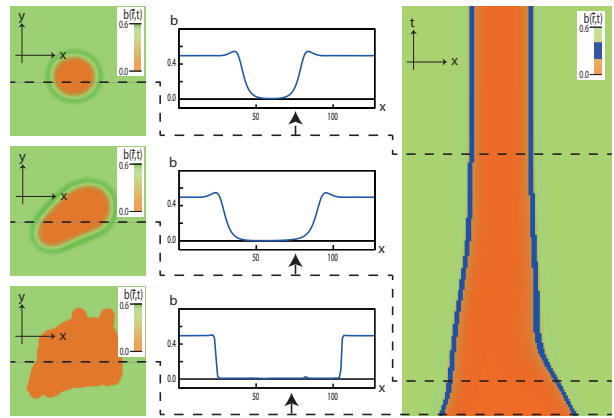}}
\caption{(right) The time evolution of non-circular shape toward the formation of FC. (middle) the cross section along the $x$ direction during time evolution; and (left) the space-time map showing the dynamics leading to the formation of FC. Same parameters as in Fig.~\ref{TW2} except $\mu=1.231$ and the mesh number integration is $128 \times 128$.}
\label{TW3}\end{figure}
Equation~(\ref{TW1}) can help to understand the behaviour of FCs, not only the stationary regime, where fronts interaction leads to the uniformly curved circular shape, but also the way that an initial perturbation evolves toward a circular shape. Suppose we start with non-uniformly curved bare state, numerical simulation of 
Eq.~(\ref{TW1}) shows that, in the course of time, the space-time dynamics  leads  to the formation of uniformly curved circular domain (see Fig.~\ref{TW3}). The circular shape results from fronts interaction mediated by Lorentzian-like non-local coupling.  The cross section along the $x$ direction shows occurrence of fringes characterised by larger biomass density. These fringes appear clearly on the natural FCs shown in Figures~\ref{TW2}a,  \ref{TW2}b. They possess a stable plateau (see the cross section along the $x$ direction Fig. 3) and the minimum biomass density is zero, corresponding to the bare state. 

\begin{figure}[t]
\centerline{\includegraphics[width=7cm]{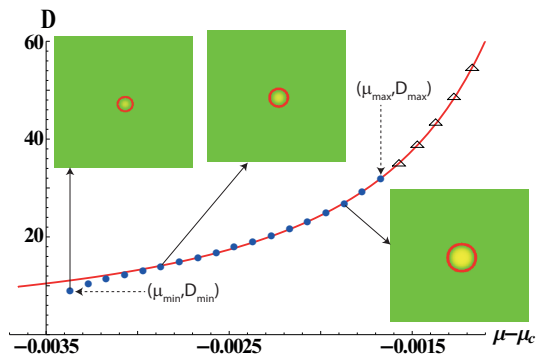}}
\caption{The diameter of the fairy circle as a function of the relative aridity with respect to the critical value $\mu_c=1.233$. Full dots indicate stable FCs, and triangles indicate unstable FCs. The parameters are the same as in Fig. 3 except the mesh number integration is $256 \times 256$. The red curve fits with the function  $(a/(\mu_c -\mu)^{\alpha})$ with $a=0.002$ and $\alpha=1.51$.}
\label{TW4}
\end{figure}

As the aridity parameter $\mu$ increases, the environment becomes more and more 
arid and the size of FCs broadens as illustrated in Fig.~\ref{TW4}. Stable FCs occur in the range of 
$\mu_{min}<\mu<\mu_{max}$. This  explains why the FCs' average diameter increases with the aridity. Indeed field measurements indicate that the FCs' average diameter varies in the range of $2$ m-$10$ m from the South to the North where the aridity increases  \cite{Rooyen}. This provides a ratio between the 
 maximum and minimum average diameter equal to $5$. The estimation by numerical 
 simulations of the diameter shows that this ratio is of the order of $4$ 
(see Fig.~\ref{TW4}).  Note, however, that the results obtained in this Letter are different from 
those generated in the pattern forming regime \cite{Tli-Lef-Vla}. Indeed, in this regime, 
the size of spots does not vary with the aridity and spots do not exhibit a  plateau. 
Therefore, the spots obtained in the pattern forming instability are not sufficient to explain 
the fairy circles. For $\mu>\mu_{max}$, FCs become unstable, 
and lose their circular shape through a curvature instability. The diameter of FCs grows, and on a long-term evolution they exhibit a radial deformation. However, when  $\mu<\mu_{min}$, the diameter of FCs decreases, so that FCs shrink and 
disappear through a transition toward the uniformly vegetated state. 
  
Two ecological   models are used to understand the formation, the maintenance and the influence of the aridity on the formation of  fairy circles. We attribute their  stabilisation to   two main ingredients. First, the ecosystem should operate in the bistability region between homogeneous covered states. Second, the strong competitive interaction between plants should be incorporated in the modelling.  Quantitative interpretation of observations and of the predictions provided by the theory are discussed. Our theoretical mechanism is  applicable to other systems with strong non-local coupling such as metamaterials, and populations dynamics \cite{Oto}.

Fruitful discussions with S. Coulibaly and R. Lefever are gratefully acknowledged.
M.T. is a Research Associate with the Fonds de la Recherche Scientifique F.R.S.-FNRS, Belgium. M.G.C. acknowledge the financial support of FONDECYT project 1120320 and C.F.O.  the financial support of   Becas Chile.

\end{document}